\documentclass[
reprint,
superscriptaddress,
nofootinbib,
amsmath,amssymb,
aps,
pra,
floatfix,longbibliography
]{revtex4-2}
\usepackage{graphicx}
\usepackage{dcolumn}
\usepackage[hypertexnames, breaklinks=true, bookmarksnumbered=true, bookmarksopen=true, colorlinks=true, linktocpage=true, citecolor=magenta, urlcolor=magenta, linkcolor=magenta]{hyperref}
\usepackage{amsmath,amssymb,amsfonts,amsthm}
\usepackage{mathtools} 
\usepackage[T1]{fontenc}
\usepackage[latin9]{inputenc}
\usepackage{txfonts}
\usepackage{bbm} 
\usepackage{bm} 
\usepackage{mathrsfs} 
\usepackage{xcolor}
\usepackage{natbib}
\allowdisplaybreaks

\newcommand{\RR}{\mathbbm{R}}

\DeclareMathOperator{\tr}{Tr}

\newcommand{\id}{\mathbbm{1}}

\newtheorem{theorem}{Theorem}

\newcommand{\bra}[1]{\langle #1|}
\newcommand{\ket}[1]{|#1 \rangle}
\newcommand{\braket}[2]{ \langle #1| #2 \rangle}
\newcommand{\ketbra}[2]{|#1 \rangle\! \langle#2|}

\newcommand{\I}{\mathscr{I}}
\newcommand{\RS}{\mathscr{R}}

\newcommand{\MI}{\hat{+}}
\newcommand{\MIi}{\hat{-}}

\begin{document}

\title{Operational Resource Theory of Imaginarity}
\author{Kang-Da Wu}
\affiliation{CAS Key Laboratory of Quantum Information, University of Science and Technology of China, \\ Hefei 230026, People's Republic of China}
\affiliation{CAS Center For Excellence in Quantum Information and Quantum Physics, University of Science and Technology of China, Hefei, 230026, People's Republic of China}

\author{Tulja Varun Kondra}
\affiliation{Centre for Quantum Optical Technologies, Centre of New Technologies, University of Warsaw, Banacha 2c, 02-097 Warsaw, Poland}

\author{Swapan Rana}
\affiliation{Centre for Quantum Optical Technologies, Centre of New Technologies,
University of Warsaw, Banacha 2c, 02-097 Warsaw, Poland}
\affiliation{S. N. Bose National Centre for Basic Sciences, JD Block, Sector III,
Kolkata 700106, India}
\affiliation{Physics \& Applied Mathematics Unit, Indian Statistical Institute, 203 B T Road,
Kolkata 700108, India}

\author{Carlo Maria Scandolo}
\affiliation{Department of Mathematics \& Statistics, University of Calgary, Calgary, AB, T2N 1N4, Canada}
\affiliation{Institute for Quantum Science and Technology, University of Calgary, Calgary, AB,  T2N 1N4, Canada}

\author{Guo-Yong Xiang}

\email{gyxiang@ustc.edu.cn}
\affiliation{CAS Key Laboratory of Quantum Information, University of Science and Technology of China, \\ Hefei 230026, People's Republic of China}
\affiliation{CAS Center For Excellence in Quantum Information and Quantum Physics, University of Science and Technology of China, Hefei, 230026, People's Republic of China}

\author{Chuan-Feng Li}

\affiliation{CAS Key Laboratory of Quantum Information, University of Science and Technology of China, \\ Hefei 230026, People's Republic of China}
\affiliation{CAS Center For Excellence in Quantum Information and Quantum Physics, University of Science and Technology of China, Hefei, 230026, People's Republic of China}

\author{Guang-Can Guo}

\affiliation{CAS Key Laboratory of Quantum Information, University of Science and Technology of China, \\ Hefei 230026, People's Republic of China}
\affiliation{CAS Center For Excellence in Quantum Information and Quantum Physics, University of Science and Technology of China, Hefei, 230026, People's Republic of China}

\author{Alexander Streltsov}
\email{a.streltsov@cent.uw.edu.pl}
\affiliation{Centre for Quantum Optical Technologies, Centre of New Technologies,
University of Warsaw, Banacha 2c, 02-097 Warsaw, Poland}

\begin{abstract} Wave-particle duality is one of the basic features of quantum mechanics, giving rise to the use of complex numbers in describing states of quantum systems, their dynamics, and interaction. Since the inception of quantum theory, it has been debated whether complex numbers are actually essential, or whether an alternative consistent formulation is possible using real numbers only. Here, we attack this long-standing problem both theoretically and experimentally, using the powerful tools of quantum resource theories. We show that -- under reasonable assumptions -- quantum states are easier to create and manipulate if they only have real elements. This gives an operational meaning to the resource theory of imaginarity. We identify and answer several important questions which include the state-conversion problem for all qubit states and all pure states of any dimension, and the approximate imaginarity distillation for all quantum states. As an application, we show that imaginarity plays a crucial role for state discrimination: there exist real quantum states which can be perfectly distinguished via local operations and classical communication, but which cannot be distinguished with any nonzero probability if one of the parties has no access to imaginarity. We confirm this phenomenon experimentally with linear optics, performing discrimination of different two-photon quantum states by local projective measurements. These results prove that complex numbers are an indispensable part of quantum mechanics.
\end{abstract}

\maketitle

\emph{Introduction}---Complex numbers, originated in mathematics, are widely used in mechanics, electrodynamics, and optics, allowing for an elegant formulation of the corresponding theory. The rise of quantum mechanics as a unified picture of waves and particles further strengthened the prominent role of complex number in physics. According to the postulates of quantum
mechanics, a state of a quantum system is described by a wave function $\Psi(x)=\left|\,\Psi(x)\,\right|\,e^{-i\,\phi(x)}$ with probability amplitude $\left|\,\Psi(x)\,\right|^2$ and phase $\phi(x)$. The wave-based point of view provides an important set of tools for the formulation and construction of quantum physics. Therefore it is natural to ask whether the complex arithmetic in quantum mechanics, arising from the imaginary part of $e^{i\,\phi}$ is necessary to describe the fundamental properties and dynamics of a quantum system. In other words, can quantum physics be restated in a formalism using real numbers only? 

One approach to address this question is to use the standard rules of
quantum mechanics, but to enforce all states and measurement
operators to have real elements only~\cite{Stuckelberg,Araki-real,Hardy+Wootters.FoP.2012,Wootters.FoP.2012,Baez,Aleksandrova+2.PRA.2013, Wootters.JPA.2014,Wootters.BC.2016,BarnumGraydonWilceCCEJA}. The aim of this approach is
then to find physical effects and applications, which are possible in standard quantum
mechanics, but impossible in its version restricted to real numbers~\citep{Wootters1990,PhysRevA.99.062110}. It has been noted that this real-vector-space quantum theory is fundamentally different from the standard one from various aspects, e.g., it is bilocally tomographic \cite{Hardy+Wootters.FoP.2012}, a rebit (real qubit) can be maximally entangled with many rebits \cite{Wootters.FoP.2012, Aleksandrova+2.PRA.2013, Wootters.JPA.2014}, and it  allows optimal transport of information from preparation to measurement~\cite{Wootters.BC.2016}.
 
Another reason to distinguish between complex and real quantum states
is the effort to establish them in experimental setups. An important example is polarization-encoded photonic system, where we can realize an arbitrary rotation around the $y$-axis by a single half-wave plate, while for a rotation around the $z$-axis two additional quarter-wave plates are needed. The fact that a certain type of transformations is easy to perform is the basic feature of any quantum resource theory~\cite{Quantum-resource-1,Quantum-resource-2,ChitambarRevModPhys.91.025001}. This justifies the study of the \emph{resource theory of imaginarity}~\citep{Hickey+Gour.JPA.2018},
using the framework of general quantum resource theories,
which has been successfully applied to investigate basic properties
and applications of quantum entanglement~\citep{HorodeckiRevModPhys.81.865},
quantum coherence~\citep{StreltsovRevModPhys.89.041003}, and quantum
thermodynamics~\citep{delRio,Lostaglio-thermo}.

The aim of this work is twofold. Firstly, we provide the resource theory of imaginarity with an operational meaning, discussing the experimental role of complex and real operations, i.e., quantum operations which do not create imaginarity. Secondly, we identify and answer several important questions within this theory. As an application, we show that imaginarity plays a crucial role for local quantum state discrimination, when complex numbers are allowed in the measurement. We show that there exist real bipartite states which
can be perfectly distinguished via local operations and classical communication (LOCC), but which cannot
be distinguished with any non-zero probability via LOCC restricted
to real local measurements. In the context of quantum tomography, a similar effect has been observed previously in~\cite{Wootters1990}. By experimentally measuring the optimal distinguishing probability for different families of mixed states, our results clearly demonstrate that complex numbers play a distinguished role in quantum theory, allowing for phenomena which would not be possible with real quantum mechanics alone.

\medskip{}

\textit{Resource theory of imaginarity}---The first step to formulating any resource theory is to identify the free states of the theory, i.e.\ quantum states which, within the theory under study, can be created at no cost. Similar to the resource theory of coherence~\cite{StreltsovRevModPhys.89.041003,BaumgratzPhysRevLett.113.140401}, we specify a particular basis $\left\{\ket{j}\right\}$, and a pure quantum state can be written as $\ket{\psi}=\sum_j c_j \ket{j}$, with complex coefficients $c_j$ which satisfy $\sum_j|c_j|^2=1$. The natural choice for free states in the theory of imaginarity
are \emph{real states}, i.e., quantum states with all coefficients $c_j$ being real (up to a non-observable overall phase)~\cite{Hickey+Gour.JPA.2018}. Mixed real states can be identified as convex combinations of real pure states $\rho =\sum_j p_j\,\ketbra{\psi_j}{\psi_j}$. The set of all real states will be denoted by $\RS$. It can also be characterized as the set of states with a real density matrix~\cite{Hickey+Gour.JPA.2018}.

The formulation of a resource theory is completed by defining an appropriate set of free operations, corresponding to physical transformations of quantum systems which are easy to implement. In general, quantum operations can be specified by a set of Kraus operators $\{K_j\}$ satisfying the completeness relation $\sum_jK_j^\dag K_j=\id$.
In the case of probabilistic transformations, the Kraus operators satisfy the more general condition $\sum_jK_j^\dag K_j\leq\id$. As the free operations of imaginarity theory we identify quantum operations which admit a Kraus decomposition having only real elements in the free basis~\cite{Hickey+Gour.JPA.2018}: $\braket{m}{K_j|n} \in \RR$ for all $j$, $m$, $n$. Such transformations are called \textit{real operations}~\cite{Hickey+Gour.JPA.2018}. This definition guarantees that real operations cannot create imaginarity, even if interpreted as a general quantum measurement. In this case, the post-measurement state will be real for any real initial state, regardless of the measurement outcome.

A desirable feature of a quantum resource theory is the existence of a golden unit: a quantum state which can be converted into any other state via free operations. In the resource theory of imaginarity the golden unit is the maximally imaginary state $\ket{\MI}=(\,\ket{0}+i\,\ket{1}\,)/\sqrt{2}$. Interestingly, via real operations it is possible to convert $\ket{\MI}$ into any state of arbitrary dimension~\cite{Hickey+Gour.JPA.2018}. Another maximally imaginary state is given by $\ket{\MIi}=(\,\ket{0}-i\,\ket{1}\,)/\sqrt{2}$. A detailed discussion of the main features of quantum resource theories, including resource quantifiers and state conversion properties under free operations, is provided in the accompanying paper~\cite{PRAversion}. If not stated otherwise, we consider quantum systems of arbitrary but finite dimension in the following.
\medskip{}

\textit{Quantum state conversion}---We will now present a complete solution for the conversion problem via real operations for all qubit states, characterizing when a qubit state $\rho$ can be converted into another qubit state $\sigma$ via real operations. To this end, recall that any single-qubit state can be represented by a real 3-dimensional Bloch vector. Now, the transition $\rho\rightarrow \sigma$ is possible via real operations if and only if 
\begin{equation}
    s_{y}^{2}\leq r_{y}^{2},\,\,\,\,\,\,\,\frac{1-s_{z}^{2}-s_{x}^{2}}{s_{y}^{2}}\geq\frac{1-r_{z}^{2}-r_{x}^{2}}{r_{y}^{2}}, \label{eq:DeterministicQubit}
\end{equation}
where $\boldsymbol{r}$ and $\boldsymbol{s}$ are the Bloch vectors of the initial and the target state, respectively. The proof of this statement is given in~\cite{PRAversion}, and relies on methods developed earlier within the resource theory of quantum coherence~\cite{StreltsovPhysRevLett.119.140402,ChitambarPhysRevLett.117.030401,ChitambarPhysRevA.94.052336}.
 
Notably, there exist states $\sigma$ which cannot be obtained from a given state $\rho$ via real operations. In this case, it might still be possible to achieve the conversion \emph{probabilistically}. Defining the conjugated state $\ket{\psi^*}=\sum_j c_j^* \ket{j}$, we now present the optimal conversion probability via real operations for any two pure states.
\begin{theorem} \label{thm:PureConversion}
The maximum probability for a pure state transformation $\ket{\psi} \rightarrow \ket{\phi}$ via real operations is given by 
\begin{equation}
    P(\,\ket{\psi} \rightarrow \ket{\phi}\,) = \min \left\{\,\frac{1-|\,\braket{\psi^*\,}{\,\psi}\,|}{1-|\,\braket{\phi^*\,}{\,\phi}\,|}, 1\,\right\}.
\end{equation}
\end{theorem}
\noindent The proof of the theorem makes use of properties of general resource quantifiers, we refer to \cite{PRAversion} for more details. We note that the theorem holds true for systems of arbitrary finite dimension. Moreover, if the target state $\ket{\phi}$ is real, the conversion probability is $1$ for any initial state.
\medskip{}

\textit{Approximate imaginarity distillation}---So far we have discussed \emph{exact} transformations between quantum states via real operations, both deterministically and stochastically. We will now go one step further, and consider \emph{approximate} transformations, in the cases when an exact transformation is impossible. Typically, one aims to convert a state $\rho$ into the most valuable quantum state, which in the resource theory of imaginarity is the maximally imaginary state $\ket{\MI}$. This leads us to the \textit{fidelity of imaginarity}, quantifying the maximal fidelity between a state $\rho$ and the maximally imaginary state, achievable via real operations $\Lambda$:
\begin{equation}
    F_\mathrm{I}(\rho)=\max_\Lambda \braket{\MI}{\Lambda[\rho]|\MI}. \label{eq:FI-2}
\end{equation}
In the following, we will provide a closed expression for $F_\mathrm{I}$, establishing at the same time a close connection to the robustness of imaginarity, defined as~\cite{Hickey+Gour.JPA.2018} 
\begin{equation}
    \I_R(\rho) =\min_\tau\left\{\,s \geq 0:\frac{\rho+s\tau}{1+s} \in \RS \,\right\},
\end{equation}
where the minimum is taken over all quantum states $\tau$ and all $s \geq 0$. We note that similar quantifiers have been studied earlier in entanglement theory~\cite{VidalPhysRevA.59.141,SteinerPhysRevA.67.054305,Plenioquant-ph/0504163} and the resource theory of coherence~\cite{NapoliPhysRevLett.116.150502,PianiPhysRevA.93.042107}.
\begin{theorem} \label{thm:FiPRL}
For any quantum state $\rho$ the fidelity of imaginarity is given as \begin{equation}\label{maxfid}
    F_\mathrm{I}\,(\,\rho\,) =\frac{1+\I_R\,(\,\rho\,)}{2} = \frac{1}{2} + \frac{1}{4}||\,\rho-\rho^T\,||_1,
\end{equation}
where $T$ denotes transposition and $||M||_1=\tr\sqrt{M^\dagger M}$ is the trace norm.
\end{theorem}
\noindent
This theorem provides at the same time a closed formula for the fidelity of imaginarity and the robustness of imaginarity for systems of arbitrary finite dimension. The proof of Theorem~\ref{thm:FiPRL} further implies that the maximum in Eq.~(\ref{eq:FI-2}) can be achieved with a real operation $\Lambda$ having a two-dimensional image space. Interestingly, these results can be extended to a general class of quantum resource theories, see Corollary 14 in~\cite{RegulaPhysRevA.101.062315}. The proof of Theorem~\ref{thm:FiPRL} is presented in~\cite{PRAversion}, where we also give more details on the robustness measure and its role in general quantum resource theories. 
\medskip{}

\textit{Applications}---We will now discuss applications of imaginarity as a resource for discrimination of quantum states and quantum channels. Channel discrimination can be seen as a game, where one has access to a ``black box'' with the promise that it implements a quantum channel $\Lambda_j$ with probability $p_j$. The goal of the game is to guess $\Lambda_j$ by applying the black box to a quantum state $\rho$, followed up by a suitably chosen positive operator valued measure (POVM) $\{M_j\}$ which satisfies $\sum_jM_j=\id,\,M_j\geq 0$. The measurement outcome $j$ then serves as a basis for guessing that the black box has implemented the channel $\Lambda_j$ in the corresponding realization of the experiment. The probability of correctly guessing an ensemble of channels $\{p_j,\,\Lambda_j\}$ in this procedure is given by $p_{\mathrm{succ}}(\,\rho,\,\{\,p_j,\,\Lambda_j\,\},\,\{M_j\} \,)=\sum_j p_j \tr [\,M_j\Lambda_j\left(\rho\right)\,]$.

Recently, it has been shown that \textit{any} quantum resource provides an operational advantage in some channel discrimination task~\cite{TakagiPhysRevLett.122.140402,Takagi+Regula.PRX.2019}. Specifically, for the resource theory of imaginarity, it holds that 
\begin{equation}
    \max_{\{p_j,\,\Lambda_j\},\,\{\,M_j\,\}} \frac{p_{\mathrm{succ}}\left(\,\rho,\,\{\,p_j,\,\Lambda_j\,\},\,\{\,M_j\,\} \,\right)}{\max_{\sigma\, \in \, \RS} p_{\mathrm{succ}}\left(\,\sigma,\,\{\,p_j,\,\Lambda_j\,\},\,\{\,M_j\,\} \,\right)} = 1+ \I_R (\rho). \label{eq:RobustnessChannelDiscrimination-1}
\end{equation}
Eq.~\eqref{eq:RobustnessChannelDiscrimination-1} implies that for any quantum state $\rho$ which has non-real elements there exists a set of channels such that the optimal guessing probability is strictly larger than for any $\sigma \in \RS$.

Another closely related task is \emph{quantum state discrimination}~\cite{chefles2000quantum}, where one aims to distinguish between quantum states $\rho_j$, each given with probability $p_j$. To this end, one performs quantum measurements, described by a POVM $\{M_j\}$. The average probability for correctly guessing the state is $p_\mathrm{succ}(\,\{p_j,\,\rho_j\},\,\{\,M_j\,\}\,)=\sum_jp_j\tr[\,M_j\rho_j\,]$. 
In general, one aims to find a strategy $\{M_j\}$ which maximizes the success probability for a given ensemble of states and probabilities $\{p_j,\rho_j\}$. For a broad class of quantum resource theories, the peformance of this task can be quantified by extending the robustness quantifier from states to measurements~\cite{Takagi+Regula.PRX.2019,UolaPhysRevLett.122.130404,Oszmaniec2019operational}.

Going one step further, we will now show that complex numbers play an indispensable role in \emph{local state discrimination} \cite{Bergia-local-tomography,Wootters1990}. Assume that the states to be discriminated are shared by two distant parties, Alice and Bob. It was shown in~\cite{walgate2000local} that any pair of pure orthogonal states can be perfectly distinguished via LOCC. To perfectly distinguish the states $\{ \rho^{AB}_j\}$ via LOCC, there must exist a POVM with elements $\{M_j\}$ of the form $M_j = \sum_k A_{j,k} \otimes B_{j,k}$ and the property $\tr (M_j\rho^{AB}_k)  = \delta_{jk}$ for all $j$ and $k$. If the states $\{ \rho^{AB}_j\}$ are real, we are particularly interested in perfect discrimination with \textit{local real operations and classical communication} (LRCC), where all $A_{j,k},\,B_{j,k}$ must be real and symmetric. Indeed, if two states are pure, orthogonal, and real, such perfect LRCC discrimination is possible, see~\cite{PRAversion} for more details.

For some real mixed states, instead, the situation is radically different. Consider the states
\begin{equation}
\begin{aligned}\rho_{1}^{AB} & =\frac{1}{2}\left(\,\ket{\phi^-}\!\bra{\phi^-}+\ket{\psi^+}\!\bra{\psi^+}\,\right),\\
\rho_{2}^{AB} & =\frac{1}{2}\left(\,\ket{\phi^+}\!\bra{\phi^+} + \ket{\psi^-}\!\bra{\psi^-}\,\right)
\end{aligned}
\label{eq:StateDiscriminationStates}
\end{equation}
with the Bell states $\ket{\phi^{\pm}}=(\ket{00}\pm\ket{11})/\sqrt{2}$, and $\ket{\psi^{\pm}}=(\ket{01}\pm\ket{10})/\sqrt{2}$. 
These states can be perfectly distinguished via LOCC. To see this, we express the states as follows: 
\begin{equation}
\begin{aligned}
    \rho^{AB}_1=\frac{1}{4}(\id+\sigma_y\otimes\sigma_y),\\
    \rho^{AB}_2=\frac{1}{4}(\id-\sigma_y\otimes\sigma_y).
\end{aligned}
\end{equation}
Consider now the POVM defined by
\begin{equation}
\begin{aligned}
    M_1 &= \ketbra{\MI}{\MI}\otimes \ketbra{\MI}{\MI} + \ketbra{\MIi}{\MIi}\otimes \ketbra{\MIi}{\MIi}, \\
    M_2 &= \ketbra{\MI}{\MI}\otimes \ketbra{\MIi}{\MIi} + \ketbra{\MIi}{\MIi}\otimes \ketbra{\MI}{\MI}.
\end{aligned}
\end{equation}
This POVM can be implemented as an LOCC protocol, if Alice and Bob perform local measurements in the $\left\{\ket{\MI},\ket{\MIi}\right\}$ basis and share their measurement outcomes via a classical channel. We verify the equality $\tr (M_j\rho^{AB}_k)  = \delta_{jk}$, implying that this POVM perfectly disctiminates the states~(\ref{eq:StateDiscriminationStates}). On the other hand, the states~\eqref{eq:StateDiscriminationStates} cannot be distinguished via LRCC with any nonzero probability. To see this, note that $\tr[S \sigma_y]=0$ for any real symmetric $2\times 2$ matrix $S$. It follows that for any POVM element $M_j = \sum_k A_{j,k} \otimes B_{j,k}$ with real symmetric matrices $A_{j,k}$ it holds 
\begin{equation}
    \tr(M_j \rho_1^{AB}) = \tr(M_j \rho_2^{AB}) =\frac{1}{4}\tr(M_j).
\end{equation}
This means that the states~(\ref{eq:StateDiscriminationStates}) are completely indistinguishable via LRCC, even if we consider imperfect state discrimination with finite error.

The states~\eqref{eq:StateDiscriminationStates} show the role of imaginarity for quantum state discrimination in an extreme way. It is clear from the above discussion that this effect is also observed if only one of the parties is limited to real operations, and the other party has access to all quantum operations locally. Nevertheless, the states can be perfectly distinguished by LOCC, if both Alice and Bob can perform general quantum measurements locally. 

These results further highlight the relevance of complex numbers in quantum mechanics. Note that the states~\eqref{eq:StateDiscriminationStates} have real elements in the computational basis. This means that they are also valid states in ``real quantum theory'' \cite{Stuckelberg,Araki-real,Hardy+Wootters.FoP.2012,Wootters.FoP.2012,Baez,Aleksandrova+2.PRA.2013, Wootters.JPA.2014,Wootters.BC.2016,BarnumGraydonWilceCCEJA}, which is the restriction of quantum theory to real states and real measurements. In such a theory, two remote parties would not be able to distinguish these states with any non-zero probability, whereas they are actually perfectly distinguishable in reality.
\medskip{}

\textit{Experimental relevance of imaginarity}---Here, we perform a comparison of real operations and general quantum operations in optical experiments, focusing on the single-photon interferometer set-up with half-(quarter-) wave plates [H(Q)WP] and polarizing beam splitters as the building blocks. Under these assumptions, a real quantum operation acting on path degree of a $d$ dimensional system can be implemented with $(d^6-d^3)/2$ unset wave plates, whereas a general quantum operation requires at least $d^6-1$ unset wave plates for the implementation. We assume that both operations are implemented via a unitary dilation. For large $d$, this implementation allows one to reduce the number of HWP(QWP) 
by $1/2$ if we restrict ourselves to real operations, see the accompanying article~\cite{PRAversion} for more details. Similar results are found in implementing a real $n$-outcome generalized measurement on a single polarization-encoded qubit~\cite{PRAversion}. Even in this case, using real measurements instead of general measurements reduces the number of unset wave plates by $1/2$ in the limit of large $n$. These results show that under assumptions commonly applied in optical experiments real states are easier to create and real operations are easier to perform when compared to general states and operations. This justifies the choice of real states (operations) as the free states (operations) of the resource theory of imaginarity~\cite{PRAversion}.  
\medskip{}

\begin{figure}
	\centering
	\includegraphics[scale=0.075]{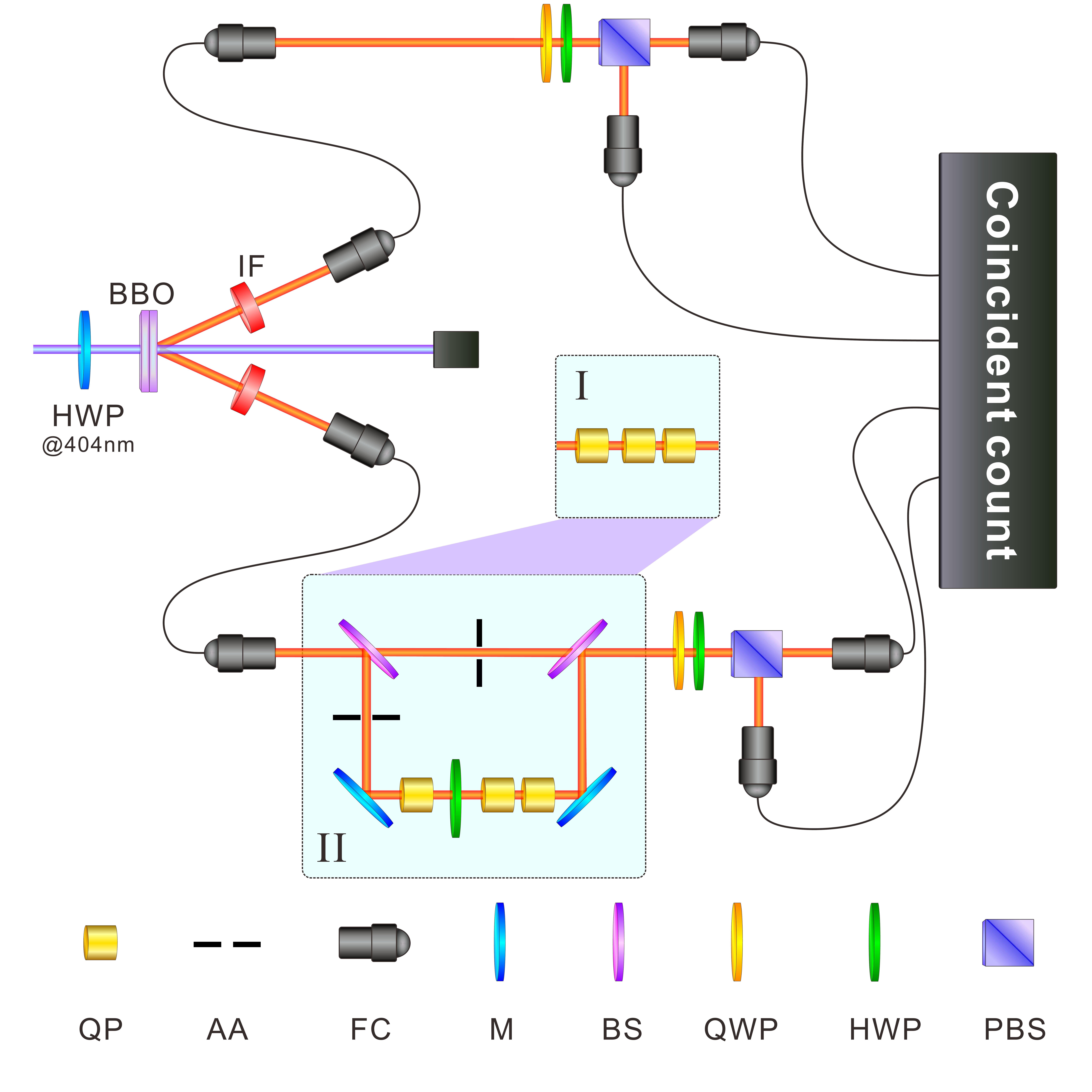}
	\caption{\label{fig:expsetup}
		\textbf{Experimental setup for local state discrimination.} The experiments are carried out using linear optics. We prepare entangled photon sources and perform local projective measurements on each photon, identifying the successful guessing probability by classical communications.
	The optical elements are: HWP, half wave plate; QWP, quarter wave plate; BBO, $\beta$-BaB$_2$O$_4$; IF, interference filter; AA, adjustable aperture; BS, beam splitter; M, mirror; QP, quartz plate; FC, fiber coupler. We refer to Ref.~\cite{note1} for more details.}
\end{figure}

\textit{Experimental local state discrimination}---As discussed earlier, imaginarity plays an important role for quantum state discrimination. We devised an experimental setup for local state discrimination using two entangled photons. The experimental setup is shown in Fig.~\ref{fig:expsetup}. After generating polarization-entangled photon pairs, we prepare different mixed states for discrimination experiments (light blue regions I and II in Fig.~\ref{fig:expsetup}). Local projective measurements are performed on each photon via the combination of a QWP, a HWP, and a polarizing beamsplitter (PBS). From the photon detecting and coincident system, we can experimentally determine the guessing probability~\cite{clarke2001experimental,mosley2006experimental,solis2017experimental}.

\begin{figure*}
	\centering
	\includegraphics[scale=0.135]{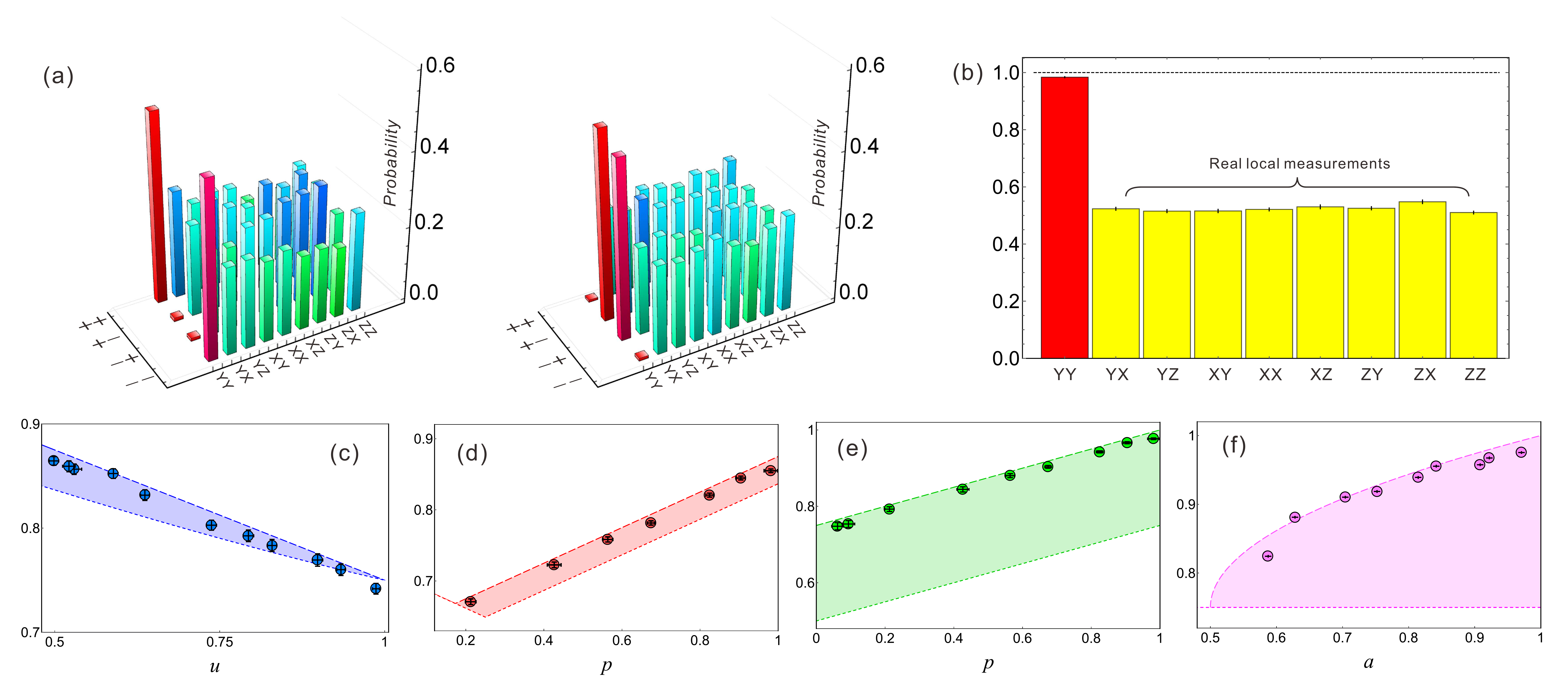}
	\caption{\label{fig:data2}
		\textbf{Experimental results for local state discrimination.} All discrimination tasks concern output states after some local free operations, when the initial state is prepared as a Bell state. (a) and (b) present the probabilities for the extreme examples in Eq.~\eqref{eq:StateDiscriminationStates}. (b) shows the experimentally measured guessing probabilities under local measurements. The horizontal axes of (a) represent the measurement results $\{++,+-,-+,--\}$, where $+,-$ corresponds to the result $\pm 1$ of local observable $A\,\in\{X,\,Y,\,Z\}$. (c-f) present results for experimental discrimination of mixed states via local projective measurements and classical communication. The families of states are prepared as follows: (c) $\ket{\phi^+}\!\bra{\phi^+}$ and $\left[\,u\ket{\phi^+}\!\bra{\phi^+}+(1-u)\ket{\psi}\!\bra{\psi}+\ket{\phi^-}\!\bra{\phi^-}\,\right]/2$, where $\ket{\psi}$ is a non trivial entangled pure state; (d) $p\ket{\phi^+}\!\bra{\phi^+}+(1-p)\id/4$ and $(\ket{\phi^+}\!\bra{\phi^+}+\ket{\psi}\!\bra{\psi}+2\ket{\phi^-}\!\bra{\phi^-})/4$; (e) $(\ket{\phi^-}\!\bra{\phi^-}+\ket{\psi^+}\!\bra{\psi^+})/2$ and $p\ket{\phi^+}\!\bra{\phi^+}+(1-p)\frac{\id}{4}$; (f) $(\ket{\phi^-}\!\bra{\phi^-}+\ket{\psi^+}\!\bra{\psi^+})/2$ and $s\ketbra{\phi^+}{\phi^+}+(1-s)\ketbra{\phi^-}{\phi^-}$, with purity $a=1-2s+2s^2$. Dashed and dotted lines represent bounds for guessing probabilities using general projective measurements and real projective measurements, respectively. The shaded areas represent the advantage of using complex measurements over real ones. In (d), general projective measurements do not provide any advantage compared to real projective measurements in the range $p < 0.173$. See~\cite{note1} for more details.
	}
\end{figure*}

Our experiment consists of two parts. In the first part we consider the discrimination of the two states in Eq.~\eqref{eq:StateDiscriminationStates}. These states can be perfectly distinguished by performing local projective measurements in the maximally imaginary basis $\left\{\ket{\MI},\ket{\MIi}\right\}$ and share the measurement outcomes via a classical channel. The experimental results are shown in Fig.~\ref{fig:data2} (a) and (b). The experimental success probability reads $P=0.984\pm0.002$. To demonstrate that the states cannot be distinguished with real local measurements, we show the experimentally measured success probabilities under the $\sigma_x$ and $\sigma_z$ measurements, since real measurements can be written as a combination of $\id$, $\sigma_x$, and $\sigma_z$. In this case the output is nearly a uniform distribution. From Fig.~\ref{fig:data2} (b) we extract experimentally determined guessing probabilities under different local Pauli measurements. We can see that imaginarity is necessary in measurement of both subsystems for improving the guessing probability.

In Fig.~\ref{fig:data2} (c-f), we show the experimental results for distinguishing different families of mixed states, given in the caption of Fig.~\ref{fig:data2}. All dashed lines represent theoretically derived maximum guessing probabilities via LOCC; dotted lines represent the aforementioned probabilities under LRCC, we refer to the caption of Fig.~\ref{fig:data2} and Ref.~\cite{note1} for more details. Note that all states considered here have only real elements in the computational basis.

These results clearly demonstrate the indispensable role of imaginarity in local state discrimination, even if the states to be distinguished have real density matrices.
\medskip{}

\emph{Discussion}---In this work, we investigate the resource theory of imaginarity, studying the role of complex numbers in quantum mechanics in an operational way. We demonstrate the usefulness of our methods in local state discrimination, where two remote parties aim to distinguish states by applying local operations and classical communication. We show -- both theoretically and experimentally -- that there exist real quantum states which can be perfectly distinguished in this setup if imaginarity is used in the local measurements. However, when restricting to only real measurements, the states cannot be distinguished with any nonzero probability. This demonstrates that complex numbers are an essential ingredient of quantum mechanics.

The usefulness of complex numbers in quantum mechanics is worth an in-depth study also in the light of the recent advances in quantum technologies. An important example is quantum computers, which can solve certain problems of interest significantly faster than any classical computer~\cite{Bravyi308,Arute2019}. As of today, the reason for this quantum advantage is not completely understood, especially when it comes to quantum computers operating on noisy states~\cite{KnillPhysRevLett.81.5672,DattaPhysRevA.72.042316,DattaPhysRevLett.100.050502,DakicPhysRevLett.105.190502,Matera_2016}. A quantitative analysis of imaginarity in quantum computers can shed new light on the quantum features required for the quantum speedup. 

\medskip{}
\textit{Acknowledgements}---We thank Bartosz Regula and Micha\l{} Oszmaniec for discussion. The work at the University of Science and Technology of China is supported by the National Key Research and Development Program of China (No. 2018YFA0306400), the National Natural Science Foundation of China (Grants No. 61905234, 11974335,11574291, and 11774334), the Key Research Program of
Frontier Sciences, CAS (Grant No. QYZDYSSW-SLH003), and the Fundamental Research Funds for the Central Universities (Grant No. WK2470000026). T.V.K., S.R., and A.S.\ acknowledge financial support by the ``Quantum Optical Technologies'' project, carried out within the International Research Agendas programme of the Foundation for Polish Science co-financed by the European Union under the European Regional Development Fund. C.M.S.\ acknowledges the hospitality of the Centre for Quantum Optical Technologies at the University of Warsaw, and financial support by the Pacific Institute for the Mathematical Sciences (PIMS) and a Faculty of Science Grand Challenge award at the University of Calgary.

\bibliography{ImaginarityBib}

\newpage

\section*{Supplemental Material}

The experimental setup contains two modules: a state preparation and a measurement module. 
In the state preparation module, we can prepare all Bell states $\ket{\phi^{\pm}}$, $\ket{\psi^{\pm}}$, Werner states $p\ketbra{\phi^+}{\phi^+}+(1-p)\frac{\id}{4}$, mixed two-qubit states $u\ketbra{\phi^+}{\phi^+}+(1-u)\ketbra{\psi}{\psi}$, and maximally correlated states $w\ketbra{\phi^+}{\phi^+}+(1-w)\ketbra{\phi^-}{\phi^-}$. For preparing the aforementioned quantum states, a maximally entangled state is firstly generated. In particular, two type I phase-matched $\beta$-barium borate (BBO) crystals, whose optic axes are normal to each other~\cite{Kwia99Ultrabright}, are pumped by the continuous semiconductor laser at 404 nm for the generation of photon pairs with a central wavelength at $\lambda$=808 nm via spontaneous parametric down conversion process (SPDC). A HWP working at $\lambda$=404 nm is set in front of the BBO crystals to control the quantum state of the generated photon pairs encoded in polarization. Fig.~1 of the main text illustrates the methods for experimentally generating Werner states, which can also be used to generate the four Bell states~\cite{wu2017experimentally,Wu2020}.

Specifically, to generate $\ket{\phi^\pm}$, we need to drop the unbalanced interferometer (UI) part (the light blue region in Fig.~1), and set the angle of the wave plate (WP@404 nm) to $50\pm22.5^\circ$. To prepare $\ket{\psi^{\pm}}$, we need an additional wave plate in one of the arms to flip one of the photonic states.

To generate Werner states
\begin{equation}
    \rho_p=p\ket{\phi^+}\bra{\phi^+}+(1-p)\frac{\id}{4},
\end{equation}
we make use of a technology named unbalanced Mach-Zehnder interferometer (the light blue region in Fig.~1), which is implemented on one of the entangled photon pairs. We use a variation of the technology presented in \cite{Wu2020,wu2017experimentally,qi2017adaptive}, with the parameter $p$ controlled by adjusting two apertures in the UI. 

To prepare the two-photon states $\frac{1}{2}(u\ket{\phi^+}\!\bra{\phi^+}+(1-u)\ket{\psi}\!\bra{\psi}+\ket{\phi^-}\!\bra{\phi^-})$ which are mixtures of three pure states, we use the ensemble average of a two-photon mixed state $u\ket{\phi^+}\bra{\phi^+}+(1-u)\ket{\psi}\bra{\psi}$ and a pure state $\ket{\phi^-}$. The preparation of $u\ket{\phi^+}\bra{\phi^+}+(1-u)\ket{\psi}\bra{\psi}$ and maximally correlated states is realized by placing adjustable quartz plates (QP) in either optical path (see Fig.~1).

We first introduce the methods for preparing the former one. The two-qubit state can be written as
\begin{equation}\label{eq:mixedstate22}
    \rho_2=p\ket{\phi^+}\bra{\phi^+}+(1-p)\tilde{O}\ket{\phi^+}\bra{\phi^+} \tilde{O}^T,
\end{equation}
where $\tilde{O}=\tilde{o}\otimes\id$ is some orthogonal matrix, $\tilde{o}=R^{-1}(\theta)\sigma_z R(\theta)$, and $R(\theta)$ is a 2-dimensional rotation.  Thus we can write Eq.~\eqref{eq:mixedstate22} as 
\begin{equation}\label{eq:mixedstate23}
    \rho_2=R^{-1}(\theta)\left[p\ket{\tilde{\phi}^+}\bra{\tilde{\phi}^+}+(1-p)\sigma_z\ket{\tilde{\phi}^+}\bra{\tilde{\phi}^+}\sigma_z\right]R(\theta),
\end{equation}
where $\ket{\tilde{\phi}^+}\bra{\tilde{\phi}^+}=R(\theta)\ket{\phi^+}\bra{\phi^+}R^{-1}(\theta)$. 

Note that $p(\cdot)+(1-p)\sigma_z\cdot\sigma_z$ is the decoherence map. Then by Eq.~\eqref{eq:mixedstate23}, it is obvious that $u\ket{\phi^+}\bra{\phi^+}+(1-u)\ket{\psi}\bra{\psi}$  can be realized by implementing a rotated local decoherence map on one subsystem. In particular, this is done by placing QPs in either path with rotation angles set to around $20^\circ$, after preparing  a maximally entangled state $\ket{\phi^+}$. We can use similar methods to prepare the maximally correlated states by placing QPs with rotation angles set to $0^\circ$.

We now focus on experimental state discrimination with local projective measurements and classical communication. The measurement module can perform all projective qubit measurements and produce corresponding coincident outcomes. In particular, we experimentally prepare the two bipartite states separately and compare the measurement outcomes. Here, we consider only projective measurements producing four outcomes experimentally.  Let us consider two projective measurements along $\boldsymbol{\alpha}$ and $\boldsymbol{\beta}$ on Alice's and Bob's system, respectively, where $\boldsymbol{\alpha}$ and $\boldsymbol{\beta}$ are directions in the Bloch ball. Recall that any bipartite state of two qubits can be written as 
\begin{equation}\label{eq:proindividual}
\rho^{AB}=\frac{1}{4}\left(\id+\sum_ja_j\sigma_j^A\!\otimes\! \id^B+\id^A\!\otimes\!\sum_jb_j\sigma_j^B+\sum_{jk}E_{jk}\sigma_j^A\!\otimes\!\sigma_k^B\right).  
\end{equation}
There are four possible outcomes $++,\,+-,\,-+,\,--$  for the two projective measurements. The probability for each outcome reads
\begin{equation}
    P_{m_a,m_b}=\frac{1}{4}\left[1+ m_a \boldsymbol{\alpha}\cdot\mathbf{a}+m_b \boldsymbol{\beta}\cdot\mathbf{b}+m_am_b\boldsymbol{\beta}^T\cdot ( E^T \boldsymbol{\alpha})\right],
\end{equation}
where $\mathbf{a}=(a_1,a_2,a_3)$, $\mathbf{b}=(b_1,b_2,b_3)$, as in Eq.~\eqref{eq:proindividual}, and $m_a=\pm1, \,m_b=\pm1$ denote measurement outcomes. The probability of distinguishing two states $\{w,\,\rho^{AB}\}$ and $\{\tilde{w},\,\tilde{\rho}^{AB}\}$ under measurements along $\boldsymbol{\alpha}$ and $\boldsymbol{\beta}$ is given by 
\begin{equation}
    p_{\mathrm{succ}}=\sum^{+,+}_{j=-,k=-}\max\left\{wP_{jk},\,\tilde{w}\tilde{P}_{jk}\right\},
\end{equation}
which can be evaluated as
\begin{equation}
    p_{\mathrm{succ}}=\frac{1}{2}\sum^{+,+}_{j=-,k=-}\left(wP_{jk}+\tilde{w}\tilde{P}_{jk}+\left|wP_{jk}-\tilde{w}\tilde{P}_{jk}\right|\right).
\end{equation}
Substituting into Eq.~\eqref{eq:proindividual} we obtain
\begin{align}
    &p_{\mathrm{succ}}=\frac{1}{2}+\frac{1}{8}\left|\boldsymbol{\alpha}\cdot\delta\mathbf{a}+\boldsymbol{\beta}\cdot\delta\mathbf{b}+\boldsymbol{\beta}^T\cdot(\Delta E^T\boldsymbol{\alpha})\right|\nonumber\\&+\frac{1}{8}\left|\boldsymbol{\alpha}\cdot\delta\mathbf{a}-\boldsymbol{\beta}\cdot\delta\mathbf{b}-\boldsymbol{\beta}^T\cdot(\Delta E^T\boldsymbol{\alpha})\right|\nonumber\\&+\frac{1}{8}\left|\boldsymbol{\alpha}\cdot\delta\mathbf{a}-\boldsymbol{\beta}\cdot\delta\mathbf{b}+\boldsymbol{\beta}^T\cdot(\Delta E^T\boldsymbol{\alpha})\right|\\\nonumber
    &+\frac{1}{8}\left|\boldsymbol{\alpha}\cdot\delta\mathbf{a}+\boldsymbol{\beta}\cdot\delta\mathbf{b}-\boldsymbol{\beta}^T\cdot(\Delta E^T\boldsymbol{\alpha})\right|
\end{align}
where $\delta\mathbf{a}=w\mathbf{a}-\tilde{w}\tilde{\mathbf{a}}$, $\delta\mathbf{b}=w\mathbf{b}-\tilde{w}\tilde{\mathbf{b}}$ and $\Delta E^T=w E^T-\tilde{w}\tilde{E}^T$.
Note that in our experiments, we have $\delta \mathbf{a}=\delta \mathbf{b}=\mathbf{0}$ theoretically. Then $p_{\mathrm{succ}}=[1+|\boldsymbol{\beta}^T\cdot(\Delta E^T\boldsymbol{\alpha})|]/2$. Thus ideally the maximum distinguishing probability is given by 
\begin{equation}
   p_{\mathrm{succ}}=\frac{1}{2}+\frac{(\Delta E)_m}{2},
\end{equation}
where $(\Delta E)_m$ is the maximum singular value of $\Delta E^T$.

For bipartite real states, we can write $\Delta E^T$ as
\begin{equation}
    \Delta E^T=\begin{pmatrix}
    &e_{xx}&0&e_{xz}\\
    &0&e_{yy}&0\\
    &e_{zx}&0&e_{zz}
    \end{pmatrix};
\end{equation}
moreover, if both Alice and Bob can only use real measurement, then $\boldsymbol{\alpha}_y=\boldsymbol{\beta}_y=0$. The maximum guessing probability is determined according to the maximum singular value of $\begin{pmatrix}&e_{xx}&e_{xz}\\&e_{zx}&e_{zz}\\\end{pmatrix}$, which we denote as $(\Delta e)_m$. When $e_{yy}>(\Delta e)_m$, any non-zero amount of imaginarity in local measurement results in an advantage over all real measurements. 

In our experiments, we measured the maximum success probabilities for four class of mixed states for the same preparation probabilities $w=\tilde{w}=0.5$. The four classes are given as follows:
\begin{align}
\mathrm{(i)}\quad & \ket{\phi}\!\bra{\phi}^{AB}=\ket{\phi^+}\!\bra{\phi^+},\nonumber\\
&\rho^{AB}_2=\frac{1}{2}\left[\,u\ket{\phi^+}\!\bra{\phi^+}+(1-u)\ket{\psi}\!\bra{\psi}+\ket{\phi^-}\!\bra{\phi^-}\,\right],\nonumber\\
\mathrm{(ii)}\quad & \rho^{AB}_1=p\ket{\phi^+}\!\bra{\phi^+}+(1-p)\frac{\id}{4},\nonumber\\
& \rho^{AB}_2=\frac{1}{4}(\,\ket{\phi^+}\!\bra{\phi^+}+\ket{\psi}\!\bra{\psi}+2\ket{\phi^-}\!\bra{\phi^-}\,),\nonumber\\
\mathrm{(iii)}\quad & \rho^{AB}_1= \frac{1}{2}(\,\ket{\phi^-}\!\bra{\phi^-}+\ket{\psi^+}\!\bra{\psi^+}\,),\\
&\rho^{AB}_2=p\ket{\phi^+}\!\bra{\phi^+}+(1-p)\frac{\id}{4},\nonumber\\
\mathrm{(iv)}\quad & \rho^{AB}_1=\frac{1}{2}(\,\ket{\phi^-}\!\bra{\phi^-}+\ket{\psi^+}\!\bra{\psi^+}\,),\nonumber\\
&\rho^{AB}_2=\frac{1}{2}(1+\sqrt{2a-1})\ket{\phi^+}\!\bra{\phi^+}+\frac{1}{2}(1-\sqrt{2a-1})\ket{\phi^-}\!\bra{\phi^-},\nonumber
\end{align} 
where $\ket{\psi}=\psi_{00}\ket{00}+\psi_{01}\ket{01}+ \psi_{10}\ket{10}+\psi_{11}\ket{11}$ is a non-trivial entangled pure state with $\psi_{00}=-\psi_{11}\approx 0.54$ and $\psi_{01}=\psi_{10}\approx 0.46$, and the parameter $a$ denotes the purity of the state $\rho^{AB}_2$.

For the states in (i), we can theoretically calculate the maximum guessing probability under local projective measurements and classical communications, which turns out to be
\begin{equation}
p_{\mathrm{succ}}=1-\frac{u}{4}.
\end{equation}
When restricted to real measurements, the maximum guessing probability is 
\begin{align}
   p_{\mathrm{succ}}'=\frac{1}{8}\left[\,6-u+\sqrt{2+2\sin\theta_0-2u\left(1+\sin\theta_0\right)+u^2}\,\right],
\end{align}
where $\theta_0=10^\circ$ ideally, corresponding to the angles of quartz plates. The guessing probability under LRCC is less than $1-u/4$. Here we used the principal value of the complex powers of $-1$.

For the states in (ii), the theoretically calculated maximum guessing probability is given by
\begin{subequations}
\begin{align}
  & p_{\mathrm{succ}}=\frac{1}{8}(2p+5) \,\,\, \mathrm{for}\,\,\, p\geq\frac{1}{8}\left(\tilde{m}-1\right), \\
  & p_{\mathrm{succ}}=\frac{9}{16}+\frac{1}{16}\left(\tilde{m}-4p\right) \,\,\, \mathrm{for} \,\,\, 0<p<\frac{1}{8}\left(\tilde{m}-1\right),
\end{align}
\end{subequations}
where $\tilde{m}=\sqrt{5+4\sin\theta_0}$, which leads to $(\tilde{m}-1)/8\approx 0.173$. When restricted to real measurements, the maximum guessing probability is given by
\begin{subequations}
\begin{align}
 & p_{\mathrm{succ}}'=\frac{1}{16}\left(7+4p+\tilde{m}\right) \,\,\, \mathrm{for} \,\,\, p\geq\frac{1}{4}, \\
  & p_{\mathrm{succ}}'=\frac{9}{16}+\frac{1}{4}\left(\tilde{m}-p\right) \,\,\, \mathrm{for} \,\,\, 0<p<\frac{1}{4}.
\end{align}
\end{subequations}
Interestingly, for $p<\left(\tilde{m}-1\right)/8$ general projective measurements do not provide any advantage compared to real projective measurements, leading to the same maximal guessing probability.

Then for  the states in (iii), maximum guessing probability is given by
\begin{equation}
    p_{\mathrm{succ}}=\frac{p}{4}+\frac{3}{4},
\end{equation}
while for real measurements, the upper bound is given by $p_{\mathrm{succ}}'=p/4+1/2$.

Finally for  the states in (iv), we can theoretically obtain the maximum guessing probability as
\begin{equation}
    p_{\mathrm{succ}}=\frac{3}{4}+\frac{\sqrt{2a-1}}{4},
\end{equation}
where $a$ represents purity. For real measurements the probability can be evaluated as $p_{\mathrm{succ}}'=3/4$.

\end{document}